\documentclass[preprint2]{emulateapj}

\usepackage{amssymb}
\usepackage{multirow}

\usepackage{graphicx}

\def\cm{\,{\rm cm}}

\def\ergscm2 {erg\,s$^{-1}$cm$^{-2}$}

\def\cm2 {cm$^{-2}$}

\def\aap {A\&A}
\def\apj {ApJ}

\shorttitle{DES13S2cmm as a signature of a QN in a He-HMXB system}
\shortauthors{Ouyed et al.}

\begin{document}
 
\title{The superluminous SN DES13S2cmm as a signature of a Quark-Nova in a He-HMXB system}
 
\author{Rachid Ouyed\thanks{Email:rouyed@ucalgary.ca}, Denis Leahy and Nico Koning}

\affil{Department of Physics \& Astronomy, University of Calgary, 2500 University Drive NW, Calgary, AB T2N 1N4, Canada}

\begin{abstract}
We show that by appealing to a Quark-Nova (the explosive transition of a neutron star to a quark star) occurring in an He-HMXB system   we can account for the lightcurve  of the first superluminous SN, DES13S2cmm, discovered by the Dark-Energy Survey.  The  neutron star's  
explosive conversion is triggered as a result of accretion during the He-HMXB second Common Envelope phase. The dense, relativistic, Quark-Nova ejecta in turn  energizes  the extended  He-rich Common Envelope in an inside-out shock heating process.
  We find an excellent fit (reduced $\chi^2$ of 1.09) to the bolometric  light-curve  of SN DES13S2cmm  including the   late time emission, which we attribute to Black Hole accretion following the conversion of the Quark Star to a Black hole.
\end{abstract}


%*********************************************************************

\section{Introduction}

DES13S2cmm is
the first confirmed  superluminous SN (SLSN) from the Dark Energy Survey
 (\cite{papadopoulos2015}).  Its bolometric light-curve (LC) shows a  slowly declining tail
 when compared to other SLSNe.   It also shows an initial hump (the first four data points
  prior to the peak; see Figure 1).
   \cite{papadopoulos2015} explored two possible power sources
 ($^{56}$Ni decay and magnetar spin-down power) but both models poorly fit the 
 LC.  In this paper we show that a Quark-Nova (QN) occurring in a  Helium-High-Mass X-ray binary (He-HMXB),  which 
experiences a second common envelope (CE) phase,    can account for the  
 LC features of SN  DES13S2cmm including the tail.  The paper is organized as follows: in \S 2 we give a brief overview of the QN model
 and the occurrence of QNe in binary systems. In \S 3 we show the results of applying the QN in a He-HMXB system
  to SN DES13S2cmm. We provide a brief discussion and some predictions in \S 4.

\section{Quark Nova (QN) Model}

The QN is the explosive  transition of a neutron star (NS) to a quark star (QS) (Ouyed et al. 2002).   The conversion energy combined with the  core collapse of the parent NS results in the ejection of $M_{\rm QN} \sim 10^{-3}$ M$_\odot$ of neutron-rich material (Ker\"anen et al. 2005; Ouyed \& Leahy 2009; Niebergal
et al. 2010). This relativistic ejecta  has a kinetic energy, $E_{\rm QN}$,   exceeding $10^{52}$ erg   and an average Lorentz factor $\Gamma_{\rm QN}\sim 10$.  The QN can occur following  the explosion
of a massive single star or in a binary system following accretion from the companion.  The key constraint is for the NS to    reach deconfinement densities in its core by becoming massive enough (e.g. via fall-back during the SN explosion
or accretion from a companion) or
via spin-down (Staff et al. 2006).  We define $M_{\rm NS, c.}$ as the
NS critical mass  above which quark deconfinement occurs in the NS triggering the QN (see  \cite{ouyed_2013a} for a recent review).  $M_{\rm NS, c.}$  varies from  $\sim 1.6M_{\odot}$ up to $2M_{\odot}$ and higher for different  equations of state (e.g. Staff et al. 2006).
However we assume $M_{\rm NS, c}=2M_{\odot}$  to account for the recent  heavy NS observed by Demorest et al. (2010); 
$\sim 2M_{\odot}$ (or heavier) quark stars can exist when one considers  interacting quarks (e.g. Alford et al. 2007).

\subsection{QNe in single-star systems: dual-shock QNe}

A dual-shock QN (dsQN) happens when the QN occurs days to weeks after the initial SN. The delay allows the QN ejecta to catch up to and collide with the SN ejecta after it has expanded to large radii (Leahy\&Ouyed 2008; Ouyed et al. 2009). 
 The SN provides the material at large radius and the QN re-energizes it, causing a re-brightening of the SN.  For  time delays 
  of days or less,   the radius of the SN ejecta is  small enough that only  a modest re-brightening results when the QN ejecta hits the SN ejecta.
   However, the neutron-rich QN ejecta lead to unique nuclear spallation signatures (Ouyed et al. 2011a). For longer  time-delays, the radius and density of the SN ejecta are such that extreme re-brightening is observed which could  explain some SLSNe \citep{ouyed_2012, kostka2014, ouyed_2013}.  For time-delays exceeding many weeks, the SN ejecta will be too large and diffuse to experience any re-brightening.  The dsQN model has been successfully applied to 
a  number of  superluminous  and double-humped supernovae (see {\it http://www.quarknova.ca/LCGallery.html} for a picture
gallery of the fits).

\subsection{Quark-Novae  in binary systems}

 A QN will more often occur in tight binaries.   To reach $M_{\rm NS, c.}$ and experience a QN  event,  the NS needs to accrete from a Roche Lobe (RL) overflowing companion  (Ouyed et al. 2011b\&c; Ouyed \& Staff 2013; see also Ouyed et al. 2015) or during 
 a CE phase of the binary  as described in this work.
   A QN in a binary provides:
 
 (i)   A means  (the relativistic QN ejecta) to shock, compress, and heat  the  NS companion
 or the CE.  In the case of a QN occurring inside a CE (which is considered in this paper), the resulting
 envelope thermal energy is given by  $E_{\rm CE, th.}= \zeta_{\rm sh.} E_{\rm QN}$ which yields an initial
 shock temperature   $T_{\rm QN, sh.}\sim   \zeta_{\rm sh.}\times  10^{9}\ {\rm K}\times  E_{\rm QN, 52}\times (M_{\odot}/M_{\rm CE}$)
 for a He-rich envelope of mass $M_{\rm CE}$ and mean molecular weight $\mu_{\rm CE}=4/3$; $E_{\rm QN, 52}$ is the QN kinetic energy in units of $10^{52}$ ergs
 while $ \zeta_{\rm sh.}$ is the shock heating efficiency.  The CE kinetic energy is 
 found from   $E_{\rm QN} = E_{\rm CE, K}+ E_{\rm CE, th.}$ where the CE thermal energy
is $E_{\rm CE, th.} = E_{\rm CE, gas} +E_{\rm CE, rad.}$, shared between the gas and the radiation.
We find that even for extreme shock efficiency $\zeta_{\rm sh.}\sim 1$ and 
for high compression ratio 
  ($\sim 4 \Gamma_{\rm QN}\sim 40$; see  Ouyed\&Staff 2013), 
   the timescale required to burn He to O (e.g. using burning rates from \cite{huang_1998}) is too large compare 
 to the adiabatic expansion timescale of the CE.

(ii) The spin-down power from the QN compact remnant (the QS), which provides an additional energy
source.   The QS is born with a magnetic field of the order of $10^{15}$ G due to 
color ferromagnetism inherent to quark matter during the transition (Iwazaki 2005).    In the scenario considered here where the QS accretes during a CE phase, the high accretion rates would lead to QS spin-up instead of spin-down and the  QS
 would  gain mass until it becomes a BH without entering a spin-down phase. Thus in the picture 
 presented here, spin-down power can be ignored.
 
(iii) The presence of a gravitational point mass (the QS)
which may slow down and trap some of the ejecta and provide accretion power  (e.g. Ouyed et al. 2011b\&c; Ouyed et al. 2014; Ouyed et al. 2015).  The conversion of the QS to a BH during the CE phase provides
  an additional source of energy  (the BH-accretion phase) which powers the LC
  in addition to energization by the QN shock.

\subsubsection{The Quark-Nova in a He-HMXB system}
\label{sec:qnia}

In previous papers we considered the detonation of a CO White Dwarf by the QN (what
we called a QN-Ia; Ouyed\&Staff 2013; Ouyed et
al. 2014; Ouyed et al. 2015).  In a QN-Ia, the system experiences a runaway mass transfer leading
to the formation of a CO-rich torus surrounding, and in the close vicinity of, the exploding NS.
In this paper we consider a QN in a  He-HMXB.
 Such a binary  is expected to form when an O/B-HMXB binary evolves through a 
CE phase of the supergiant stellar component.  If the binary survives this  first CE event, the resultant system would contain a He-rich core (Hall\& Tout 2014)  with a NS in a relatively tight orbit. 
 For a large mass ratio\footnote{E.g. $q > 3.5$ or orbital period $P_{\rm orb.} < 0.1$ days (Ivanova et al. 2003). 
  Alternate conditions leading to a second CE phase give $2.4 \le q \le 2.7$ and $P_{\rm orb.} < 0.25$ days (Dewi \& Pols 2003). See Tauris et al. (2015) for other outcomes and for a recent investigation of the evolution of
these systems.} $q$, a runaway mass transfer is expected once the He-rich core overflows its
RL  leading to the onset of a second CE phase\footnote{In Ouyed et al. (2015), where the NS companion
is a degenerate WD,  the runaway accretion is due to companion expansion (since $R\propto M^{-1/3}$) which
leads to high-density (degenerate) torus formation.}.
 The NS accretes from two mass reservoirs (the first CE and the second CE).  For a NS
born with an initial mass of $\sim 1.6M_{\odot}$,  accretion of $\sim 0.4M_{\odot}$  during the two CE
phases is necessary to drive it above $M_{\rm NS, c.}$;  accreting  $\sim 0.2M_{\odot}$ 
of material per CE phase is not unreasonable and should be enough  to eject the first CE (\cite{brown_1995,armitage_2000}). 
If the in-spiralling NS  accretes   enough mass  during the second CE  phase, it should drive the NS above 
above $M_{\rm NS, c.}$ to undergo a QN explosion in an extended CE;
 the CE expands radially outward as the NS in-spirals towards the core. 
 Following the QN explosion the QS continues to in-spiral into the core
  while gaining mass. Below we show that if the QS turns into a BH,
  BH-accretion can power the slowly declining tail of DES13S2cmm.
 A combination of QN shock heating and BH-accretion
 provides the best fit to DES13S2cmm.

\begin{table*}[t!]
\label{table:params}
\begin{center}
\caption{Best-fit parameters  for the DES13S2cmm LC  in our model.}
%\scalebox{0.8}{
\begin{tabular}{|c||c|c||c|c|c||c|c|c|c||c|} \hline
  & \multicolumn{2}{c||}{He-rich (i.e. second) CE} &  \multicolumn{3}{|c||}{QN} & \multicolumn{4}{|c|}{BH Accretion} & \multicolumn{1}{||c|}{Fit} \\
  \hline
  &  $M_{\rm CE}$ ($M_{\odot}$) & $R_{\rm CE, 0} (R_{\odot})$    &   $v_{\rm QN}$ (km/s) &  $T_{\rm QN, sh.}$ (K) &  $\alpha_{\rm QN, sh.}$ & $t_{\rm d}$ (days) & $y_0$  & $L_{\rm 0}$ (erg/s)  & $n$ & $\chi_{\rm red.}^2$ \\\hline
BH-accretion & --  & --  & --  & --  & -- &  13.7 & 0.10  &  $6\times 10^{44}$   & 0.7 & 2.77 \\\hline
QN$+$BH-accretion &  2  & 1350  &  40,000  & $3.7\times 10^6$ & 3/2& --$^{\dagger}$ & 0.45$^{\dagger\dagger}$  &  $2.8\times 10^{44}$   & 0.8 & 1.09\\\hline
\end{tabular}\\
%}\\
$^{\dagger}$ Here, $t_{\rm d}$ is given by $t_{\rm d} = (2/3)\sqrt{ M_{\rm CE} \kappa_{\rm th}/(0.3\beta c v_{\rm QN})} =12.6$ days.
$^{\dagger\dagger}$This corresponds to $t_0 = y_0 t_{\rm d}\simeq 5.6$ days.\\
The  kinetic energy of the CE/ejecta after the QN is $E_{\rm CE, K}= (1/2)M_{\rm CE}v_{\rm QN}^2\simeq 3.2\times 10^{52}$ ergs.
~\\~\\~\\
 \end{center}
\end{table*}

\section{Application to DES13S2cmm}

   The free parameters in our model are (see Table 1):   the  CE mass
($M_{\rm CE}$), its initial (extended) radius $R_{\rm CE, 0}$  at the time of  QN shock breakout;  the CE expansion velocity
$v_{\rm QN}$ induced by the QN shock
and   QN shock heating per particle $(3/2) k_{\rm B}T_{\rm QN, sh.}$; $\alpha_{\rm QN, sh.}$, the
shock propagation parameter (see Appendix). There are three parameters for the BH-accretion model
 based on  the prescription of  \cite{dexter_2013} (see their appendix A): $L_0$ (erg s$^{-1}$),
  $y_0$ and   $n$ which together define  the injection power  $L(t)=L_0 (t/t_0)^{-n}$ with
$t_0=y_0 t_{\rm d}$ and  $t_{\rm d}$  the photon diffusion time in the CE;  BH-accretion turns on
at $t=t_0$ so that $L(t)=0$ for $t< t_0$.
Table 1 shows the model's best-fit parameters  for the BH-accretion model alone and
 the for the two-component QN$+$BH-accretion model.  The best-fits reduced $\chi^2$ values are given in the last column
 of the table.

 \subsection{BH-accretion fit}

 This model does not have the QN explosion if  the NS  experiences a transition to a BH directly while the system is  in the second CE phase.    The NS could turn into a BH prior to or  during merging
 with the CE core which forms  an accretion disk around the BH  to 
 power the LC.  The  best-fit is obtained  for  $t_{\rm d}=13.7$ days,
 $y_0=0.1$ (i.e. $t_0=5.6$ days), $L_0=6\times 10^{44}$ erg s$^{-1}$ and $n=0.7$. 
  The  fit    is shown in Figure \ref{fig:fig1} along-side the observations from \cite{papadopoulos2015}.  
The  resulting best-fit reduced $\chi_{\rm red.}^2$  is  $\sim 2.77$. The initial
hump is not well fit with this model.

\subsection{QN$+$BH-accretion fit}

\begin{figure}[h!]
\centering
\includegraphics[scale=0.7]{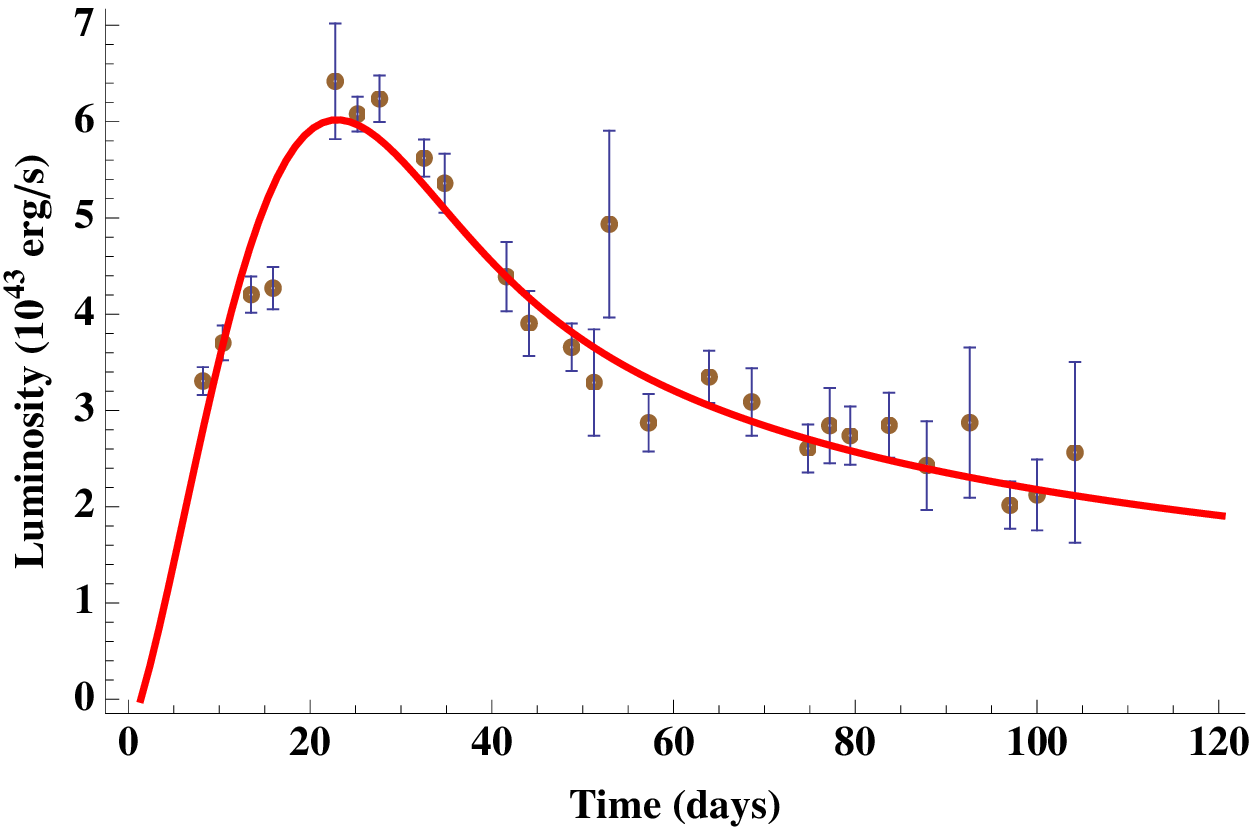}
\caption{The BH-accretion model fit (solid line) to the light curve of SN DES13S2cmm.  The observations (the solid circles
and the error bars) are from \cite{papadopoulos2015}.}
\label{fig:fig1}
\includegraphics[scale=0.7]{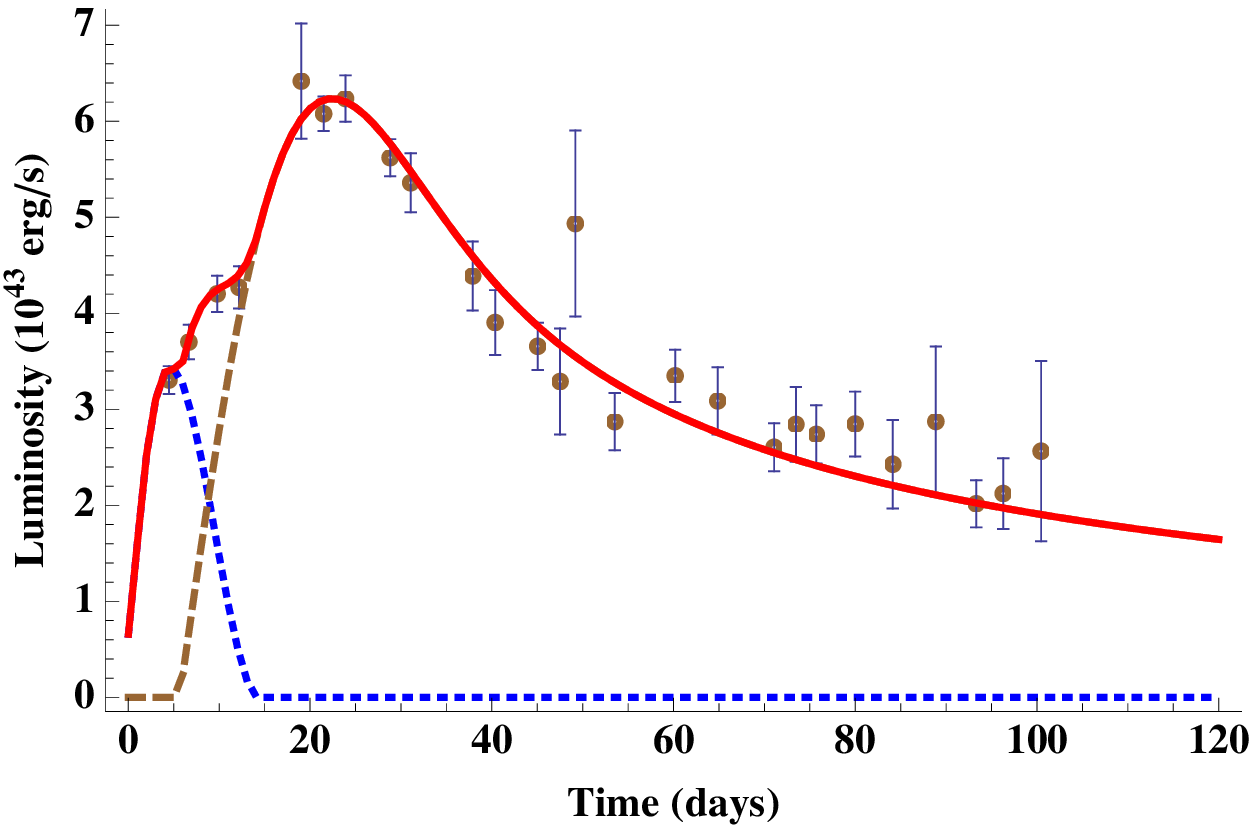}
\caption{The QN$+$BH-accretion model fit (solid line) to the light curve of SN DES13S2cmm.} 
\label{fig:fig2}
\end{figure}

Adding the QN shock heating yields a significant improvement in the LC fit with 
 a reduced $\chi_{\rm red.}^2$ of $\sim 1.09$ (see Table 1).   
 The BH-accretion phase  is delayed from the QN event by the time required for the QS to turn into a BH,
merge with the core and trigger accretion. This time delay is $t_{\rm 0}$ at which point the CE has extended to a radius
 $R_{\rm CE, 0}+ v_{\rm QN} t_0$;  $R_{\rm CE, 0}$ is the CE radius at QN shock breakout (see Appendix).   The LC fit of DES13S2cmm   is shown in Figure \ref{fig:fig2} along-side the observations from \cite{papadopoulos2015}.  

 The QN shock  energizes the He CE (of mass
 $M_{\rm CE}=2M_{\odot}$ and radius $R_{\rm CE, 0}=1350R_{\odot}$)
 and yields the initial bright and short-lived hump; the corresponding
initial CE temperature $T_{\rm CE, 0}$ is calculated from the QN shock heating ($T_{\rm QN, sh.} = 3.7\times 10^6$ K;
i.e. $\zeta_{\rm sh.}\sim 10^{-2}$ for $M_{\rm CE}=2M_{\odot}$)
by including radiation energy density (see Appendix). The photon diffusion timescale is 
$t_{\rm d} = (2/3)\sqrt{ M_{\rm CE} \kappa_{\rm Th.}/(0.3\beta c v_{\rm QN})}=12.6$ days with $\beta=13.8$ 
 (\cite{arnett_1982}); $c$ is the speed of light and $\kappa_{\rm Th.}$ the Thompson cross-section.

\section{Discussion and Conclusion}

    In this paper we showed that the LC of SN DES13S2cmm is best fit with the QN$+$BH-accretion model.
  In our model, the QN occurs during the second CE phase of He-HMXB system followed
  by a BH-accretion phase after the QS turns into a BH in the core of the CE.  The QN shock
  re-energizes the extended CE explaining the initial hump while BH-accretion power nicely
  fits the slowly declining tail. No nuclear burning is triggered by the QN shock which means that 
  there should be little or no He-burning products in SN DES13S2cmm
   and similar explosions.

   The lack of He-HMXB systems when compared to the  observed large population of HMXBs with Be-type stars (e.g. \cite{raguzova2005} and reference therein) has been used as an  observational argument in favor of mergers during the dynamically unstable mass transfer phase of the progenitors (e.g. Linden et al. 2012; see also van den Heuvel 1976). However, even in extreme cases, the theoretically
    predicted fraction of He-HMXBs surviving the CE phase is still  higher than the observed number \citep{linden2012}.
     If the NS is born massive, it will likely accrete enough matter during the first CE phase to reach $M_{\rm NS, c.}$.
     The resulting QN could remove enough matter to unbind the system and bypass the production of He-HMXBs.
      We thus speculate that the QN may be partly responsible  for the rarity of  He-HMXBs.

    If the NS does not accrete enough mass to go QN in the second CE phase,
 it may reach the center
 and form a Thorne-$\dot{Z}$ytkow object (\cite{thorne1977}). However,  continued accretion 
 should  trigger a QN leading to the same outcome. I.e. Thorne-$\dot{Z}$ytkow objects
 could be short-lived.  On the other hand, extreme accretion rates could turn
  the NS directly into a BH.

   The idea of a QN  in binaries
has proven  successful in accounting for some features of SNe-Ia (Ouyed et al. 2014; Ouyed et al. 2015) and Gamma Ray Bursters (Ouyed et al. 2011c)
 and has been able to account for the   LC of DES13S2cmm as shown here. 
Its ability in fitting  properties of unusual SNe  (see  {\it http://www.quarknova.ca/LCGallery.html}) 
  suggest that QNe may be  an integral part of binary evolution; the QN  could 
  lead to novel and interesting evolutionary paths. 
    Nevertheless,  as we have stated before, our model relies on the feasibility of the QN explosion
 which requires sophisticated simulations of the burning of a NS to a QS which are being pursued. 
 Preliminary  simulations  with consistent treatment of nuclear and neutrino reactions,
diffusion,  and hydrodynamics show instabilities that
could lead to a detonation (Niebergal et al. 2010; see also
Herzog \& R\"opke (2011) and  \cite{Manrique_2015}).  We have already argued that a ``core-collapse" QN could also
result from the collapse of the  quark matter core
(Ouyed et al. 2013a) which provides another avenue for the QN explosion.

\begin{acknowledgements}   

  We thank the referee for comments that helped improve this paper. We also thank Thomas Tauris
for valuable discussion. This work is funded by the Natural Sciences and Engineering Research Council of Canada. 

\end{acknowledgements}

%---------------------------------------------------------------

%---------------------------------------------------------------
\begin{appendix}

Due to the outward diffusion of photons, the photosphere is moving inward in mass coordinates,
  slowly at first but faster as the density decreases in time. The ejecta
   interior to the photosphere we refer to as the core.
 We will assume that the thermal energy in the exposed mass in the photosphere 
  (as the cooling front  creeps inward) is promptly radiated. The interplay between 
  uniform expansion and radiation diffusion defines the evolution of the photosphere as
  \begin{equation}
 R_{\rm phot.} (t) = R_{\rm CE}(t) - D(t)\ ,
 \end{equation}
 where $R_{\rm CE}(t)=R_{\rm CE, 0}+ v_{\rm QN}t$ and $R_{\rm CE, 0}$  the CE envelope radius 
at QN shock breakout (which corresponds to $t=0$ in our model). 
 Here $D(t)$ is the diffusion length  
 \begin{equation}
 D(t)^2 = D_0^2 + \frac{c}{n_{\rm CE}(t)\sigma_{\rm Th.}} t\ ,
 \end{equation}
 where   $n_{\rm CE}(t)= N_{\rm CE}/V_{\rm CE}(t)$ is
 the number density in the CE. The total  number of particles in the CE is
   $N_{\rm CE} = (M_{\rm CE}/\mu_{\rm CE} m_{\rm H})$ while $V_{\rm CE}(t) = (4\pi/3) R_{\rm CE}(t)^3$
    is the volume extended by the CE  and $m_{\rm H}$  the Hydrogen atomic mass. 
We define $D_0$ as the initial diffusion length scale
 by setting $n_{\rm CE, 0}\sigma_{\rm Th.} D_0\simeq 1$
  where $n_{\rm CE, 0}= N_{\rm CE}/V_{\rm CE, 0}$ and the initial
  volume $V_{\rm CE, 0} = (4\pi/3) R_{\rm CE, 0}^3$.

The initial QN shock heating per particle $(3/2)k_{\rm B} T_{\rm QN, sh.}$ is a free parameter; $k_{\rm B}$ is the Boltzmann constant. 
The heat is redistributed  between gas and radiation to get the post-shock CE temperature $T_{\rm CE, 0}$. The
relevant equation is
$\frac{3}{2} k_{\rm B} n_{\rm CE, 0}  T_{\rm CE, 0}  + a_{\rm rad} T_{\rm CE, 0}^4 =  \frac{3}{2} k_{\rm B} n_{\rm CE, 0}  T_{\rm QN, sh.}$, 
with $n_{\rm CE, 0}$ the  number  density  (of electrons and ions) and $a_{\rm rad}$ the radiation constant.

 The subsequent evolution of the CE core temperature after the CE is fully shocked is given by
 $T_{\rm core} (t) = T_{\rm CE,0} (R_{\rm CE, 0}/R_{\rm CE}(t))^{2-\alpha_{\rm QN, sh.}}$.
  To account for a non-uniform initial temperature, we
  introduce   $\alpha_{\rm QN, sh.}$ which parameterizes complex shock physics beyond the scope of this work. 
   $\alpha_{\rm QN, sh.}=0$ corresponds to the case of an adiabatic expansion with spatially uniform initial $T_{\rm CE, 0}$.
   I.e. the internal energy includes only gas internal energy with $\gamma=5/3$, so as time increases, $T_{\rm core} (t) \propto R_{\rm CE}(t)^{-2}$.    $\alpha_{\rm QN, sh.}$ also allows to account for the presence of radiation,
   e.g. for spatially uniform $T_{\rm CE, 0}$ and $\gamma=4/3$, one uses $\alpha_{\rm QN, sh.}=1$.
     $\alpha_{\rm QN, sh.}>0$ also can correspond  to a radially decreasing initial CE temperature so that
     $T_{\rm core}$ decreases more slowly than it would be for uniform $T_{\rm CE, 0}$.

 The corresponding luminosity is
\begin{equation}
\label{eq:luminosity} 
L_{\rm QN}(t)  =  c_{\rm V, tot.} (t) \Delta T_{\rm core} (t) n_{\rm  CE} (t) 4\pi R_{\rm phot.}(t)^2 \frac{d D(t)}{d t}\ ,
\end{equation}  
where the  total specific heat is 
$c_{\rm V, tot.} (t) =   c_{\rm V, gas} + c_{\rm V, rad.} = \frac{3}{2} k_{\rm B}+ \frac{a_{\rm rad} T_{\rm core} (t)^3}{n_{\rm CE} (t)}$. 
  Here, $\Delta T_{\rm core}\sim T_{\rm core}$ since the photosphere cools promptly (i.e.
  cooling time is much less than the diffusion timescale). The inward
  photospheric velocity in mass coordinates is  $d D(t)/dt$. 
  When $n_{\rm CE}(t)$ is low, the $T^3$ term dominates   so that the temperature is much lower than in the pure gas model. As $n_{\rm CE}(t)$ increases, the gas energy density becomes more important and $T$ rises.

\end{appendix}
%\clearpage

\end{document}